# Composition in the Function-Behaviour-Structure Framework

*Bob Diertens*

section Theory of Computer Science, Faculty of Science, University of Amsterdam

*ABSTRACT*

We introduce composition in the function-behaviour-structure framework for design, as described by John Gero, in order to deal with complexity. We do this by connecting the frameworks for the design of several models, in which one is constrained by the others. The result is a framework for the design of an object that supports modularity. This framework can easily be extended for the design of an object with more than one layer of modularity.

## 1. Introduction

In software engineering, dealing with complexity is a major issue and it is the ground for many software development methodologies. Most of these methodologies however do not take into account the nature and process of design. Each methodology has its success stories, but one can seldom relate it to a more abstract framework for design. A well-known method for dealing with complexity in other engineering disciplines is modelling. By making a model one can leave out some detail and concentrate on the bigger picture. Even a model can be too complicated. In previous work [1] we introduced refinement to deal with this complexity. In this article we deal with complexity through composition. This results in a design consisting of several components that interact with each other, and in which each component has a separate framework for design.

Although modularity is applied in software engineering, the problem remains that the design is largely focused on a too low level of abstraction. This is caused by the fact that software is build cheaply, and can be done over and over again. This makes it possible to test on the lowest level and often results in a race to the lowest level to start testing early in the design process. Instead of introducing modularity on a high level, modules are introduced on a low level first. Development of software is done with a focus on building on these low level modules. In that process, the higher level design is discarded and complexity is taken into the lower levels instead of dealing with it on the higher levels of design.

In our view it is better to incorporate methodologies that follow the nature and process of design, and start on a high level. An important factor in this is to know what design really is. John Gero has described a general framework for design [2] that is based on function, behaviour, and structure of the object to be designed. This framework, however, omits composition explicitly. For a thorough understanding and execution of the design process it is better to make composition explicit in the design process.

In section 2 we give an overview of the function-behaviour-structure framework for design. We introduce composition in this framework in section 3 in order to support modularity explicitly.

## 2. The Function-Behaviour-Structure Framework

In [2] Gero describes a framework for design that has sufficient power to capture the nature of the concepts that support design processes. This framework, that involves the relation between function, behaviour, and structure of a design, can be applied to any engineering discipline. Together with Kannengiesser, Gero describes the framework in [3] in relation with the environment in which designing takes place, accounting for the dynamic character of the context. We give an overview of the elements and processes that form the function-behaviour-structure (FBS) framework.

The FBS framework elements has the following elements.

|  |  |
|---|---|
| function ($F$) | The set of functions expressing the requirements and objectives that must be realized by the object. |
| structure ($S$) | Describes the components of the object and their relationships. |
| expected behaviour ($B_e$) | The set of expected behaviours to fulfill the function $F$. |
| structure behaviour ($B_s$) | The set of behaviours the structure $S$ exhibits. |
| description ($D$) | The description of the design, giving all the information to build the object, and what more there is to know about the design. |

These elements are connected in the framework by processes (Figure 1).

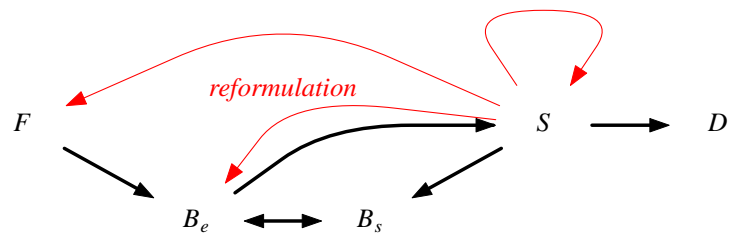

**Figure 1.** The FBS framework

An outline of the process of the FBS framework is given below.

|  |  |
|---|---|
| formulation ($F \rightarrow B_e$) | Transforming the function $F$ into behaviour that is expected from the object. |
| synthesis ($B_e \rightarrow S$) | Transforming the expected behaviour into a solution intended to exhibit this behaviour. |
| analysis ($S \rightarrow B_s$) | Deriving of the actual behaviour from the synthesized structure. |
| evaluation ($B_e \leftrightarrow B_s$) | Comparing the behaviour derived from the structure with the expected behaviour. |
| documentation ($S \rightarrow D$) | Producing the design description for the constructing or manufacturing of the object. |

In addition the framework contains reformulation processes that are carried out, based on the outcome of the evaluation of behaviours.

structure reformulation ($S \rightarrow S$)
: Changing of the structure in order to obtain a behaviour that is more in line with the expected behaviour.

behaviour reformulation ($S \rightarrow B_e$)
: Adjusting of the expected behaviour that fits the required function and is more in line with the behaviour of the structure.

function reformulation ($S \rightarrow F$)
: Changing of the function due to a better insight in the problem.

## 3. Composition of FBS frameworks

To capture modularity in design (modular design) we compose a framework ($C$–$FBS$) out of several FBS frameworks. We consider the design of the models $M$, $M^1$ and $M^2$, where the latter two are models for components of model $M$. All models have their own design process, $FBS$, $FBS^1$, and $FBS^2$, each of which can be described by the functions-behaviour-structure framework for design, see Figure 2.

In the figure the relations between the frameworks are indicated. The function of the components is determined by the description for the model $M$, and the structures for the components are to be part of the structure for model $M$.

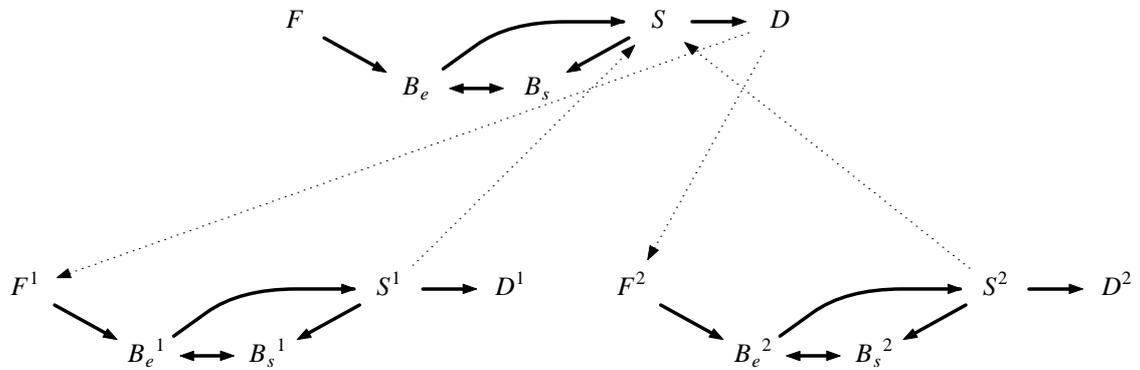

**Figure 2.** Relations between design frameworks for three models

We like to integrate the three design processes so that the processes that play a role in the composition become clear and that immidiate feedback can take place between the particular frameworks. In the following sections we describe the processes that integrate the three frameworks into one.

*3.1 Decomposition*

Once an acceptable structure $S$ is determined, the design for several of the components can be done separately. For the design of the components, the functions for each of the components have to be determined.

    function decomposition ($\{F, D\} \rightarrow F^i$)

        As the structure $S$ consists of the components and their interaction for model $M$, the description $D$ contains the functionality for the components. Furthermore, $F$ may contain functionality not contained in $D$, but that is to be taken into account in the design of the components.

This decomposition of the functions is indicated in Figure 3.

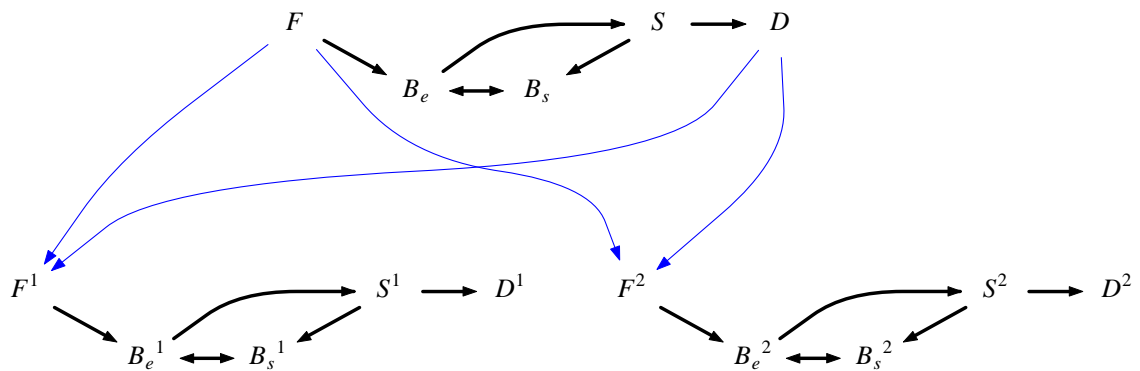

**Figure 3.** Decomposition processes in the design framework

## 3.2 Reformulation

Each *FBS* framework in the *C−FBS* framework contains the normal reformulation processes. However, reformulations in the frameworks $FBS^1$ and $FBS^2$ have to be such that the elements stay in accordance with the function *F* and description *D* of the *FBS* framework. When that is not possible anymore, the design for the component has to be rejected and one of the following reformulation processes has to take place.

> structure reformulation ($S^i \to S$)
>> When reformulation of $S^i$ is not possible anymore to obtain an acceptable structure for the component, reformulation of *S* has to take place.
>
> function reformulation ($F^i \to F$)
>> When a reformulation of the part of $F^i$ that originates from *F* is necessary, this has to be done through reformulation of *F* directly in order to keep a consistent description of the functionality throughout the *C−FBS* framework.

This reformulation processes are indicated in Figure 4.

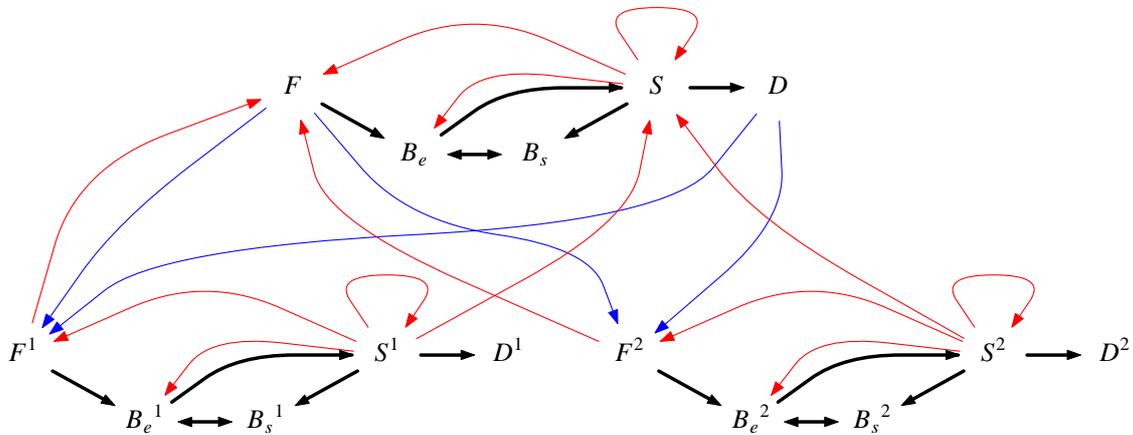

**Figure 4.** Reformulation processes in the design framework

## 3.3 Integratation

Once the design of a component is complete it has to be integrated with the overall design. The following processes describe the integration of designs.

> documentation integration ($\{D^i\} \to D$)
>> The description for each of the components is integrated with the description for the whole object.

These integration processes are indicated in Figure 5.

## 4. Conclusions

We introduced composition in the FBS framework by connecting frameworks for the design of several models. The resulting composite framework can be used for the further decomposition of the design framework, resulting in more levels of modularity in the design. We can turn the composite framework into the original framework by considering the decomposition processes as reformulations and abstract from the details of the decomposition processes. In the composition framework the modularity in the design is made explicit.

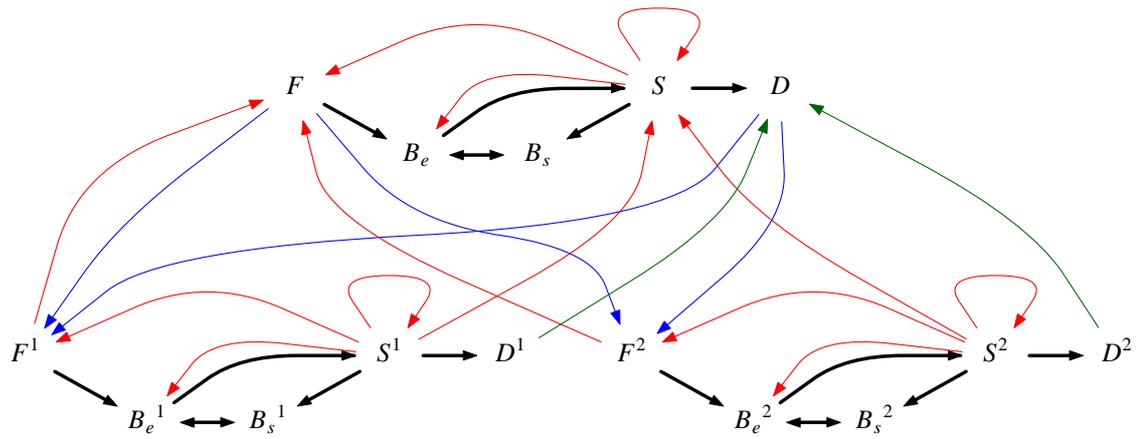

**Figure 5.** Integration processes in the design framework

## Acknowledgements

Many thanks to Alban ponse for his proofreading and feedback.